\documentclass[epj]{svjour}

\usepackage{graphics}

\begin{document}
\authorrunning{B. Saghai {\it et al.}}
\titlerunning{Search for missing baryon resonances...}
\title{Search for missing baryon resonances via associated strangeness 
photoproduction}
\author{B. Saghai\inst{1}, J.-C. David\inst{1}, B. Juli\'a-D\'{\i}az\inst{2}, 
and T.-S.H. Lee\inst{3}
}                     
\institute{Laboratoire de
recherche sur les lois fondamentales de l'Univers, DAPNIA/SPhN, 
CEA/Saclay, Gif-sur-Yvette, France 
\and ECM, Facultat de Fisica, Universitat de Barcelona, E-08028 Barcelona, Spain 
\and Physics Division, Argonne National Laboratory, Argonne, IL 60439, USA
}
\date{Received: date / Revised version: date}
\abstract{Differential cross-section and single polarization observables in
the process $\gamma p~\rightarrow~K^+ \Lambda$ are investigated within a 
constituent quark model and a dynamical coupled-channel formalism. The effects of
two new nucleon resonances and of the K*(892)- and K1(1270)-exchanges 
are briefly presented. 
} 
\maketitle
\section{Introduction}
\label{intro}
Photoproduction of mesons appears to be a promising area to investigate issues
related to the missing baryon resonances, predicted by different QCD-inspired 
approaches~\cite{MR}. Recent works, summarized e.g. in Ref.~\cite{CC06}, show
indications on few of them. In that latter publication, we
have reported on significant contributions from new $S_{11}$ and $D_{13}$
resonances to the process $\gamma p~\rightarrow~K^+ \Lambda$
studied in the total center-of-mass energy range 
W $\equiv \sqrt {s}$ $\approx$ 1.6 GeV to 2.6 GeV, 
which corresponds to the baryon resonances mass region. 

In this short report, we focus on the interplay between {\it s-} and {\it t-}channel
contributions and the duality hypothesis.
%
%
\section{Theoretical frame and {\it t-}channel issues}
\label{sec:1}

  In order to study the kaon photoproduction on the proton, 
  we have developed~\cite{CC06,CC04} multistep coupled channel formalisms for the reactions
 $\pi N~\rightarrow~\pi N$,
 $\pi N~\rightarrow~K Y$,
 $K Y~\rightarrow~K Y$, and
 $\gamma p~\rightarrow~K Y$.
  The $\pi N~\rightarrow~\pi N$ potential comes from an advanced version of effective 
  Lagrangians approaches~\cite{SH} using a unitary transformation method. 
  The same method is also used to derive from effective Lagrangians 
  the basic non-resonant $\pi N~\rightarrow~K Y$ and
 $K Y~\rightarrow~K Y$ transition potentials. 
  The direct channel $\gamma p~\rightarrow~K^+ \Lambda$ is handled within a 
  chiral constituent 
  quark model~\cite{CC06,SL} based on the $SU(6)\otimes O(3)$ broken symmetry, 
  with the starting point being the low energy QCD Lagrangian~\cite{MG}.
The four components for the photoproduction of
pseudoscalar mesons based on the QCD Lagrangian are:
\begin{eqnarray}\label{eq:Mfi}
{\cal M}_{fi}&=&{\cal M}_{seagull}+{\cal M}_s+{\cal M}_u+{\cal M}_t
\end{eqnarray}
The first term in Eq.~(\ref{eq:Mfi}) is a seagull term. It is generated by the gauge 
transformation of the axial vector $A_{\mu}$ in the QCD Lagrangian.
The second and the third terms correspond to the {\it s-} and {\it u-}channels,
respectively. 
The last term is the {\it t-}channel contribution and contains two parts: 
{\it i)} K$^+$ exchange;  
{\it ii)} K*- and K1-exchanges.
In our previous investigations~\cite{CC06,CC01}, the dynamics of our models 
  were partially based on the duality
hypothesis, according to which,
at any given energy one could use {\it either} the {\it s}-channel resonance
prescription {\it or} the {\it t}-channel Regge pole description, provided
that we sum over an {\it infinite} number of terms. An approximation
to this idea and widely discussed in the literature~\cite{Duality}, 
is to express the physical scattering 
as a series of {\it s}-channel resonances plus a general background.
In those works, we had adopted this approximation, given that our approach 
allows us to take into account individual contributions
from all known nucleon resonances in the first and second resonance regions, 
and treat as degenerate higher mass resonances. 
However, with the advent of data at high energies (W $\ge$ 2.4 GeV), it is desirable to
find out how well higher mass resonances are handled and if there is
need to include {\it t-}channel resonances. 
If so, we ought to study the drawback of such treatment on the missing
resonances issues.
%
%
\section{Results and discussion}
\label{sec:2}
We have extended the formalism presented in Ref.~\cite{CC06} to embody
the {\it t-}channel  K*- and K1-exchanges~\cite{ELA-SL}.

The fitted data base contains 1029 data points released recently: 
differential cross-sections from SAPHIR~\cite{ELSA}, 
recoil-$\Lambda$ polarization from JLab~\cite{JLab04} 
and GRAAL~\cite{GRAAL}, as well as polarized beam asymmetry 
from GRAAL~\cite{GRAAL}. 
Those data span the following angular and energy ranges
in the phase space: 
18$^\circ$ $\le$ $\theta_K^{c.m.}$ $\le$ 162$^\circ$, 
0.912 GeV $\le$ $E_{\gamma}^{lab}$ $\le$ 2.575 GeV, corresponding to the 
center-of-mass energy range 1.6 GeV $\le$ W $\le$ 2.4 GeV.
\begin{table}
\caption{Reduced $\chi^2$s for the reaction mechanism configurations, 
as explained in the text.}
\label{tab:1}       
\begin{center}
\begin{tabular}{cccc}
\hline\noalign{\smallskip}
Configuration & a & b & c \\
\noalign{\smallskip}\hline\noalign{\smallskip}
$\chi^2_{d.o.f}$ & 1.83 & 3.48 & 5.25 \\
\noalign{\smallskip}\hline
\end{tabular}
\end{center} 
\end{table}
At this stage of our study, differential cross-section data from 
 JLab~\cite{JLab06} and LEPS~\cite{LEPS06} have not been included in the 
fitted data base in order to avoid possible confusion between {\it t-}channel effects 
and consequences of inconsistencies among different data base, 
as, for example, discussed in Refs.~\cite{CC06,Mart06}. 
\begin{figure} [bh]
\resizebox{0.45\textwidth}{!}{\includegraphics*{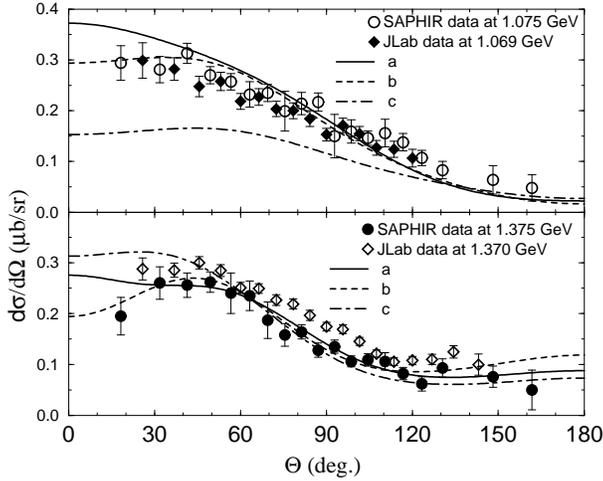}
}
\caption{Differential cross-section for the reaction 
$\gamma p \to K^+ \Lambda$ 
as a function of the outgoing
kaon angle in the center-of-mass frame.
The curves are: a) full model (full curves), b) full model but with {\it t-}channel K* and K1
contributions switched-off (dashed curves), c) full model, but with contributions from
the new 3$^{rd}$ $S_{11}$ and $D_{13}$ resonances swicthed-off (dotted-dashed curves).
The curves in the upper box are for $E_{\gamma} ^{lab}$ = 1.075 GeV and in the lower one
for $E_{\gamma} ^{lab}$ = 1.375 GeV.
Data are from SAPHIR~\cite{ELSA} and JLab~\cite{JLab06}.
}
\protect\label{fig:1}      
\end{figure}
%
 
In order to make clear the respective roles played by {\it t-}channel contributions and the 
new resonances, we have performed minimizations for three configurations with respect 
to the reaction mechanism, namely,
{\bf a)} full model: it includes all known nucleon and hyperon resonances, 
{\it t-}channel contributions from the exchange of K*(892) and K1(1270), 
as well as contributions from new $S_{11}$ and $D_{13}$ resonances. 
In the subsequent two configurations, the following contributions
have been {\it swicthed-off}:
{\bf b)} K*- and K1-exchanges; {\bf c)} new resonances.

\begin{figure}
\resizebox{0.45\textwidth}{!}{\includegraphics*{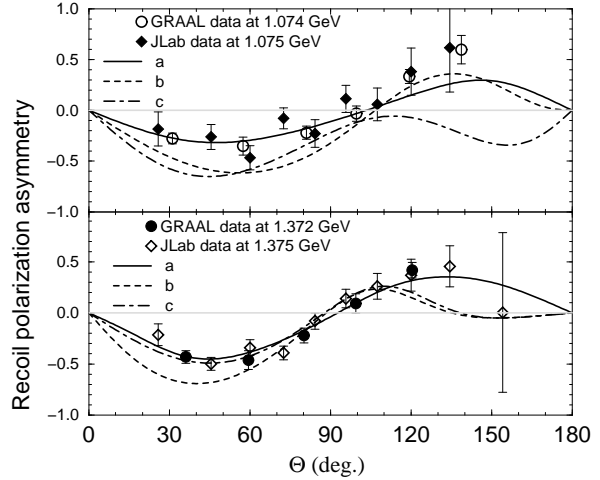}
}
\caption{Same as Fig.~\ref{fig:1}, but for recoil $\Lambda$ 
polarization asymmetry.
Data are from GRAAL~\cite{GRAAL} and JLab~\cite{JLab04}.
}
\label{fig:2}      
\end{figure}

\begin{figure}
\resizebox{0.45\textwidth}{!}{\includegraphics*{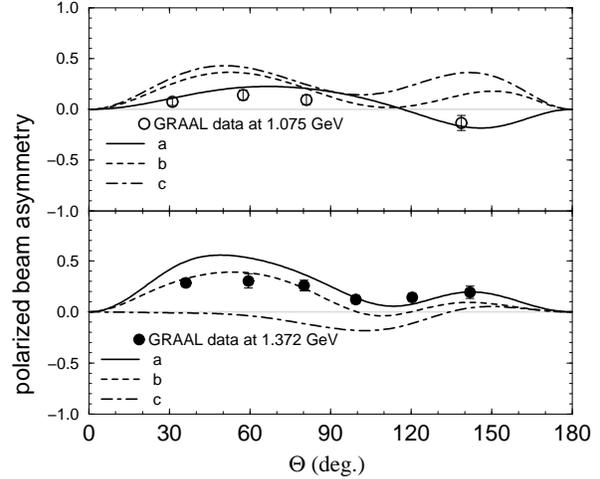}
}
\caption{Same as Fig.~\ref{fig:1}, but for  polarized beam asymmetry.
Data are from GRAAL~\cite{GRAAL}.}
\label{fig:3}      
\end{figure}

In Figs.~\ref{fig:1}-\ref{fig:3} the results of those three configurations are 
compared with the data around $E_{\gamma}^{lab}$ = 1.075 GeV and 1.375 GeV, 
which correspond to 
the total center-of-mass energy W $\approx$ 1.7 and 1.9 GeV, respectively.
Because of lack of space, here we single out results at only two energies, 
which are in the mass region of the new resonances. This choice is also
related to the fact that the GRAAL~\cite{GRAAL} data are limited to 
$E_{\gamma} ^{lab}~\le$ 1.5 GeV.
%
%
%
%
%

Results for the differential cross-section of the reaction 
$\gamma p~\rightarrow~K^+ \Lambda$ are depicted in 
Fig.~\ref{fig:1}, where the JLab data~\cite{JLab06} (not fitted) are
also shown. Figures ~\ref{fig:2} and~\ref{fig:3} embody results for
single polarization asymmetries for recoil-hyperon polarization
($\gamma p~\rightarrow~K^+ {\overrightarrow {\Lambda}}$) and 
polarized beam (${\overrightarrow {\gamma}} p~\rightarrow~K^+ \Lambda$).
\begin{table*}
\caption{Sensitivity of the investigated observables to the contributions
from the {\it t-}channel ({\cal TC}) K*- and K1-exchanges, and from the two new 
resonances ({\cal NR}), corresponding to the configurations b) and c), as
explained in the text. The most significant angular regions per energy
are given in columns 3 to 6.
}
\label{tab:2}       
\begin{center}
\begin{tabular}{lcccccc}
\hline\noalign{\smallskip}
\multicolumn{1}{c}{} &\multicolumn{1}{c}{} & 
\multicolumn{2}{c}{$E_{\gamma}^{lab}$ = 1.075 GeV }  
&\multicolumn{1}{c}{} &
\multicolumn{2}{c}{$E_{\gamma}^{lab}$ = 1.375 GeV} \\
\noalign{\smallskip}\hline\noalign{\smallskip}
Observable &Switched-off & forward  & backward  &&
forward & backward \\
& & hemisphere & hemisphere &&
hemisphere & hemisphere  \\
\noalign{\smallskip}\hline\noalign{\smallskip}
a) Cross-section &{\cal TC}& $\le$ 40$^\circ$ & - && $\le$ 30$^\circ$ & -  \\
                 &{\cal NR}& 30$^\circ$ to 90$^\circ$ & - &&
                  $\le$ 40$^\circ$ & - \\[5pt]
b) Recoil asymmetry &{\cal TC}& 30$^\circ$ to 90$^\circ$ & - && 
                      20$^\circ$ to 50$^\circ$ & -  \\
                 &{\cal NR}& 30$^\circ$ to 70$^\circ$ & $\ge$ 130$^\circ$ &&
                  - & $\ge$ 130$^\circ$ \\[5pt]
c) Beam asymmetry  &{\cal TC}& 30$^\circ$ to 60$^\circ$ & 130$^\circ$ to 150$^\circ$ 
                  && - & -  \\
                   &{\cal NR}& 30$^\circ$ to 60$^\circ$ & 130$^\circ$ to 150$^\circ$ 
                  && 30$^\circ$ to 90$^\circ$ & 90$^\circ$ to 120$^\circ$  \\
\noalign{\smallskip}\hline
\end{tabular}
\end{center} 
\end{table*}

The full model (full curves) leads to a $\chi ^2_{d.o.f}$ of 1.83
(Table~\ref{tab:1}), and the 
extracted mass and width for the two new resonances are: 
$S_{11}$($M$ = 1.820 GeV, $\Gamma$ = 240 MeV)
and $D_{13}$($M$ = 1.920 GeV, $\Gamma$ = 160 MeV). The full model gives a
reasonable account of the whole fitted data base.

Removing the {\it t-}channel resonances (dashed curves) and refitting
the data, increases the  
$\chi^2_{d.o.f}$ by almost a factor of 2 (Table~\ref{tab:1}).
In Fig.~\ref{fig:1}, the extreme forward angle data are better reproduced
in the absence of {\it t-}channel contributions. This is not the case for data
at higher energies (W $\ge$ 2 GeV), as expected from the duality hypothesis.
Finally, minimization within the configuration {\bf c)}, which embody 
no new resonances,
leads to a deterioration of the $\chi^2_{d.o.f}$ by about a factor of 3
(Table~\ref{tab:1}).
Switching-off the {\it t}-channel resonances or the new nucleon resonances produces
significant effects on the investigated observables in various angular regions,
depending on the incident photon energy. Those features appearing in 
Figs.~\ref{fig:1}-\ref{fig:3} are summarized in Table~\ref{tab:2}.
%
%
\section{Conclusions}
\label{conclu}
Among the three observables reported here, {\it t}-channel 
contributions show the largest 
effects in the recoil-$\Lambda$ polarization. The new resonances introduced
here produce noticeable features on all three observables in a rather broad range of
phase space. Those features are not mimicked by the {\it t-}channel resonances studied
here, showing that the {\it s-}channel resonances embodied in our approach
satisfy to a large extent the duality requirements. 
The extracted Mass and width values for those resonances are
$S_{11}$($M$ = 1.820 GeV, $\Gamma$ = 240 MeV)
and $D_{13}$($M$ = 1.920 GeV, $\Gamma$ = 160 MeV). Those values are compatible 
with findings from other works, namely, 
for $S_{11}$ Refs.~\cite{CC06,SL,S11} and for
$D_{13}$ Refs.~\cite{CC06,Mart06,D13}. 
%
%
\section{Acknowledgments}
\label{Thanks}
We are indebted to Frank Tabakin for his contribution to this work
at earlier stages and for enlightening discussions.
We wish to thank the GRAAL collaboration and especially,
Annick Lleres and Dominique Rebreyend for 
having provided us with their data~\cite{GRAAL} prior to publication. 
%
%


\begin{thebibliography}{}
%
\bibitem{MR} See e.g. 
	S. Capstick and W. Roberts,
        Prog. Part. Nucl. Phys. \textbf{45}, (2000) 5241; and references therein.
%
\bibitem{CC06} B. Juli\'a-D\'{\i}az, B. Saghai, T.-S.H. Lee, F. Tabakin, 
Phys. Rev. C \textbf{73}, (2006) 055204.
%
\bibitem{CC04} W.-T. Chiang, B. Saghai, F. Tabakin, T.-S.H. Lee, 
Phys. Rev. C \textbf{69}, (2004) 065208.
%
\bibitem{SH} T. Sato and T.-S.H. Lee,
Phys. Rev. C \textbf{54}, (1996) 2660; {\it ibid} C \textbf{63}, (2001) 055201;
A. Matsuyama, T. Sato, T.-S.H. Lee, arXiv nucl-th/0608015.
%
\bibitem{SL} B. Saghai and Z. Li, 
Eur. Phys. J. A \textbf{11}, (2001) 217.
%
\bibitem{MG} A. Manohar and H. Georgi, 
	 Nucl. Phys. B \textbf{234}, (1984) 189;
	Z. Li, H. Ye, M. Lu, 
        Phys. Rev. C \textbf{56}, (1997) {1099}.
%
\bibitem{CC01} W.-T. Chiang, B. Saghai, F. Tabakin, T.-S.H. Lee, 
Phys. Lett. B \textbf{517}, (2001) 101.
%
\bibitem{Duality}  See e.g.
	P.~Collins, 
	\textit{An Introduction to Regge Theory and High Energy Physics},  
	(Cambridge University Press, Cambridge,1977).
%
\bibitem{ELA-SL}
%
        J.C. David, C. Fayard, G. H. Lamot, B. Saghai,
        Phys. Rev. C \textbf{53}, (1996) 2613.
%
\bibitem{ELSA} The SAPHIR collaboration, K.H. Glander {\it et al.}, 
Eur. Phys. J. A \textbf{19}, (2004) 251.
%
\bibitem{JLab04} The CLAS Collaboration, J.W.C. McNabb {\it et al.}, 
Phys. Rev. C \textbf{69}, (2004) 042201.
%
\bibitem{GRAAL} The GRAAL Collaboration, A. Lleres {\it et al.},
accepted for publication in Eur. Phys. J. A; D. Rebreyend, 
private communication (2006).
%
\bibitem{JLab06} The CLAS Collaboration, R. Bradford {\it et al.}, 
Phys. Rev. C \textbf{73}, (2006) 035202.
%
\bibitem{LEPS06}The LEPS Collaboration, R.G.T. Zegers {\it et al.}, 
Phys. Rev. Lett. \textbf{91}, (2003) 092001;  
The LEPS Collaboration, M. Sumihama {\it et al.}, 
Phys. Rev. C \textbf{73}, (2006) 035214.
%
\bibitem{Mart06} 
         T. Mart and A. Sulaksono,
         arXiv: nucl-th/0609077.
\bibitem{S11}
        Z. Li and R. Workman,
        Phys. Rev. C \textbf{53}, (1996) R549;
       A. \v{S}varc and S. Ceci, arXiv: nucl-th/0009024;
       G.-Y Chen {\it et al.},
        Nucl. Phys. A \textbf{723}, (2003) 447;
       B. Saghai and Z. Li,
       \textit{Proceedings of NSTAR 2002 Workshop on the Physics of Excited Nucleons},
       Pittsburgh, PA (USA), 2002; Editors S.A. Dytman and E.S. Swanson (World
       Scientific, New Jersey, 2003),
       arXiv: nucl-th/0305004.
%
\bibitem{D13}
N.G. Kelkar, M. Nowakowski, K.P. Khemchandani, S.R. Jain,
           Nucl. Phys. A \textbf{730}, (2004) 121;
  A.V.~Anisovich {\it et al.},
  Eur. Phys. J. A \textbf{25}, (2005) 427;
%
  A.V.~Sarantsev {\it et al.},
  Eur. Phys. J. A \textbf{25}, (2005) 441.
%
\end{thebibliography}
\end{document}